\documentclass[aps,prl,twocolumn,amsmath,amssymb,superscriptaddress,nobibnotes,showpacs]{revtex4}

\usepackage{graphicx}
\usepackage{dcolumn}
\usepackage{bm}
\usepackage{amsmath}
\usepackage{amssymb,amsthm}
\usepackage{epsfig,color}
\usepackage[sans]{dsfont}

\definecolor{nblue}{rgb}{0.2,0.2,0.7}
\definecolor{ngreen}{rgb}{0.2,0.6,0.2}
\definecolor{nred}{rgb}{0.7,0.2,0.2}
\definecolor{nblack}{rgb}{0,0,0}

\newcommand{\tr}{\text{tr}}

\newcommand{\Id}{\openone}

\def\E{\mathcal{E}}

\def\tr{\mbox{tr}}
\def\bea{\begin{eqnarray}}
\def\eea{\end{eqnarray}}

\begin{document}

\title{Experimental implementation of the universal transpose operation using structural physical approximation}

\author{Hyang-Tag Lim}
\email{forestht@gmail.com}
\affiliation{Department of Physics, Pohang University of Science and Technology (POSTECH), Pohang, 790-784, Korea}

\author{Young-Sik Ra}
\affiliation{Department of Physics, Pohang University of Science and Technology (POSTECH), Pohang, 790-784, Korea}

\author{Yong-Su Kim}
\affiliation{Department of Physics, Pohang University of Science and Technology (POSTECH), Pohang, 790-784, Korea}

\author{Joonwoo Bae}
\email{bae.joonwoo@gmail.com}
\affiliation{School of Computational Sciences, Korea Institute for Advanced Study, Seoul, 130-012, Korea}

\author{Yoon-Ho Kim}
\email{yoonho72@gmail.com}
\affiliation{Department of Physics, Pohang University of Science and Technology (POSTECH), Pohang, 790-784, Korea}

\date{\today}

\begin{abstract}
The universal transpose of quantum states is an anti-unitary transformation that is not allowed in quantum theory. In this work, we investigate approximating the universal transpose of quantum states of two-level systems (qubits) using the method known as the structural physical approximation to positive maps. We also report its experimental implementation in linear optics. The scheme is optimal in that the maximal fidelity is attained and also practical as measurement and preparation of quantum states that are experimentally feasible within current technologies are solely applied.
\end{abstract}

\pacs{03.65.Ud, 03.67.Bg, 42.50.Ex}

\maketitle


The postulate of quantum theory that global dynamics of given quantum systems must be unitary is the constraint given to legitimate operations on quantum states. The general class of quantum operations is then found in the reduced dynamics of
subsystems and mathematically characterized by completely positive
(linear) maps over Hilbert spaces \cite{ref:Kraus}. In particular,
positive but not complete positive maps, simply called positive
throughout, are of unique importance in quantum information theory
that, for any entangled state, there exists a positive map that
determines whether or not the given state is entangled \cite{ref:Horodecki}.

The fact that there are impossible operations in quantum theory leads in a natural way to the problem of building optimal approximate quantum operations. Moreover, impossible operations in quantum theory are in general not only of fundamental and theoretical interest to characterize computational and information-theoretic capabilities of quantum information processing, but also of practical importance in implementation of approximate quantum operations for applications.

Systematic approximation to positive maps, known as the structural physical
approximation (SPA), has been proposed in Ref.~\cite{ref:HorodeckiEkert} in the context of detecting entanglement of unknown quantum states, i.e. even before identifying
given quantum states through the state tomography. The initial
proposal of the SPA assumed collective measurement for spectrum estimation that requires, i) quantum memory that stores quantum states in the quantum level for a while, and ii) coherent quantum operations that allows general manipulation of copies of
quantum states \cite{ref:Keyl}. Interestingly, however, it was shown recently that the
SPA to some positive maps correspond to quantum measurement
\cite{ref:fiu}. Furthermore, it has been recently conjectured that SPAs to
positive maps are in general entanglement-breaking, meaning that the
actual implementation of the SPA would be much simpler than originally proposed \cite{ref:Korbicz}. Note that entanglement-breaking channels can be constructed with measurement and preparation of quantum states \cite{ref:entbreak}. The conjecture has been extensively tested with known examples of positive maps, without a counter-example, for instance, in Refs.~\cite{ref:Korbicz, ref:conjex}.

In this work, we report linear optical implementation of approximating the transpose operation for a two-dimensional quantum system (i.e., photonic qubit) using the SPA based on measurement and preparation of quantum states. Our SPA scheme for the universal transpose operation is optimal in that the maximum fidelity is attained and is also practical as only single-copy level measurement and preparation of quantum states are
applied. To the best of our knowledge, our work is the first proof-of-principle demonstration that shows SPAs to positive maps are experimentally feasible within the present-day technology. 


The motivation behind the transpose operation being particularly chosen to demonstrate the SPA based on measurement and preparation of quantum states is twofold. On one hand, in the side of further applications, transpose to a subsystem is the well-known
criteria that efficiently detects useful entanglement (i.e., entangled states having negative eigenvalues after the partial transpose) \cite{ref:Horodecki,ref:Peres}. Since any approximate map that can experimentally detect entanglement via the SPA can always be factorized into a convex combination of another SPAs to non-physical operations for individual systems \cite{ref:spalocc}, our work immediately implies the feasibility of
experimental implementation of the entanglement detection via SPAs. On the other hand, in the fundamental point of view, the transpose represents the anti-unitarity in the symmetry transformation in quantum theory \cite{ref:wigner}. Since any anti-unitary is composed of a unitary and the transpose, the transpose is the only symmetry transformation
that is not allowed in quantum theory.



Let us first briefly describe the theory behind the SPA based on measurement and preparation of quantum states \cite{ref:Korbicz}.
The SPA to the transpose $T$ of a $d$-dimensional quantum state $\sigma$ in general works by admixing the complete contraction $D[\sigma] = \tr[\sigma] \Id / d$ to the positive map,
\bea
T \longrightarrow \widetilde{T} = (1-p) T + p D, \nonumber
\eea
such that the resulting map $\widetilde{T}$ is completely positive. The minimal $p$ that brings the complete positivity is known as $p=d/(d+1)$. From the well-known isomorphism between states and channels in Ref.~\cite{ref:Jamiolkowski}, the channel $\widetilde{T}$ corresponds to the so-called Jamiolkowski state $\rho_{\widetilde{T}} = [\Id\otimes\widetilde{T}](|\phi^{+}\rangle\langle\phi^{+}|)$ where $|\phi^{+}\rangle = \sum_{i}|ii\rangle/\sqrt{d}$. If the state $\rho_{\widetilde{T}}$ is separable, then the channel $\widetilde{T}$ is entanglement-breaking, meaning that the channel can be described by measurement and preparation of quantum states \cite{ref:entbreak}.


Consider now the approximate transpose of a qubit state $\rho$. In this case, the Jamiolkowski state is $\rho_{\widetilde{T}} = (|\phi^{+}\rangle \langle \phi^{+}| + |\phi^{-}\rangle \langle \phi^{-}| + |\psi^{+}\rangle \langle \psi^{+}|)/3$, where $|\phi^{-}\rangle = (|00\rangle - |11\rangle)/\sqrt{2}$ and $|\psi^{+}\rangle = (|01\rangle + |10\rangle)/\sqrt{2}$. The state is separable, having the following separable decomposition \bea \rho_{\widetilde{T}} = \frac{1}{4} \sum_{k = 1}^{4} |v_k \rangle\langle v_k | \otimes | v_k \rangle \langle v_k |, \label{eq:jamistate} \eea
where the vectors $|v_{i}\rangle$ are normalized and given by,
\bea \left| {v_1 } \right\rangle  & \propto & {\left| 0 \right\rangle  + \frac{{ie^{i\pi 2/3} }}{{i + e^{ - i\pi 2/3} }}\left| 1 \right\rangle }, ~
\left| {v_2 } \right\rangle  \propto  {\left| 0 \right\rangle  - \frac{{ie^{i\pi 2/3} }}{{i - e^{ - i\pi 2/3} }}\left| 1 \right\rangle}, \nonumber\\
\left| {v_3 } \right\rangle  & \propto &  {\left| 0 \right\rangle  + \frac{{ie^{i\pi 2/3} }}{{i - e^{ - i\pi 2/3} }}\left| 1 \right\rangle }, ~
\left| {v_4 } \right\rangle  \propto {\left| 0 \right\rangle  - \frac{{ie^{i\pi 2/3} }}{{i + e^{ - i\pi 2/3} }}\left| 1 \right\rangle }.\nonumber \eea

\begin{figure}[t]
\includegraphics[width=3in]{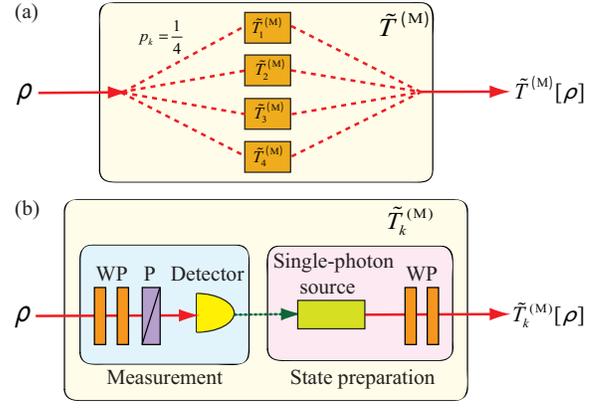}
\caption{ (a) The scheme in eq.~(\ref{eq:spat}) is shown. For an input qubit state $\rho$, an operation $\widetilde{T}^{(M)}_k$ for $k=1,2,3,4$ is randomly applied, so that the state results in the (classical) equal mixture of the four possibilities.
(b) Each $\widetilde{T}^{(M)}_k$ for $k=1,2,3,4$ denotes the operation composed by measurement and state preparation. In our experimental setup, $\rho$ is  a polarization state of a single photon. For the measurement and preparation of states, waveplates (WP) and polarizers (P) are used. The detector and the single-photon source in the measurement and the state preparation stages, respectively, become superfluous in our experiment and thus were not implemented. See text for details.}\label{fig:setup}
\end{figure}

Using the Jamiolkowski isomorphism \cite{ref:Jamiolkowski}, the SPA to the transpose of a given qubit state $\rho$ can be expressed in terms of the above vectors as,
\bea \widetilde T^{(M)}\left[ \rho  \right] = \sum\limits_{k = 1}^4
{{\rm{tr}}\left[ \frac{1}{2}  {\left| {v_k^* } \right\rangle
\left\langle {v_k^* } \right|\rho} \right]\left| {v_k }
\right\rangle \left\langle {v_k } \right|}, \label{eq:spat}\eea
where the superscript $(M)$ denotes that the scheme is measurement-based and $| v_k^* \rangle$ is complex conjugate of  $| v_k \rangle$. The set of positive operators $\{  |v_{k}^{*}\rangle \langle v_{k}^{*}|/2 \}$ defines a properly normalized measurement due to the trace preserving property of the channel $\widetilde{T}^{(M)}$. Then, eq.~\eqref{eq:spat} can be interpreted as carrying out the approximate transpose of $\rho$ in two steps: i) measuring the state $\rho$ in the basis $|v_{k}^{*}\rangle$ with equal probabilities for $k=1,2,3,4$, and ii) depending on the measurement outcome, preparing the corresponding state $|v_{k}\rangle$. The schematic diagram of eq.~(\ref{eq:spat}) is shown in Fig.~\ref{fig:setup} where the two-step operation is denoted as $\widetilde{T}_{k}^{(M)}$ for $k=1,2,3,4$.

The maximum fidelity that can be achieved by the SPA to the transpose operation, $\widetilde T^{(M)}$, can be calculated by considering a pure qubit state $|\psi\rangle$ and is given as,
\bea F = \tr[T[|\psi\rangle \langle \psi|] ~\widetilde{T}^{(M)}[|\psi\rangle\langle\psi|] ] = 2/3 \approx 0.666. \label{eq:2/3} \eea
It should be immediately clear from the above result that our scheme is optimal in that maximum fidelity of $2/3$ can be achieved \cite{ref:optimal}.

Let us now describe the linear optical realization of the SPA to the transpose shown in eq.~(\ref{eq:spat}). To prepare a single-photon polarization qubit $\rho$, we made use of the heralded single-photon source based on spontaneous parametric down-conversion (SPDC). A 2 mm thick type-II BBO crystal was pumped by a 405 nm diode laser operating at 100 mW, producing a pair of orthogonally polarized 810 nm photon pairs. Conditioned on the detection of the vertically polarized trigger photon, the horizontally polarized signal photon is prepared in the single-photon state (i.e., heralded single-photon state). The polarization state of the heralded single-photon can then be transformed to an arbitrary state by using a set of half- and quarter-wave plates, i.e., the single-photon polarization qubit \cite{Kim}. Using 10 nm FWHM  bandpass filters in front of both the trigger and the signal detectors, we observed the coincidence rate around 4 kHz.


To implement the approximate transpose operation $\widetilde{T}^{(M)}$, as shown in eq.~(\ref{eq:spat}), measurement and preparation of the single-photon polarization qubit in four different settings (randomly chosen with equal probability) are required. The ideal approximate transpose operation $\widetilde{T}^{(M)}$ is therefore composed of four $\widetilde{T}^{(M)}_k$ ($k=1,2,3,4$) and each $\widetilde{T}^{(M)}_k$ operation (randomly chosen with equal probability) consists of particular measurement and state preparation, see Fig.~\ref{fig:setup}. For each $\widetilde{T}^{(M)}_k$, the measurement in one of the basis $|v^{*}_{k}\rangle$  is performed by setting waveplates in such a way that the incoming photonic polarization qubit $\rho_{\rm{in}}$ passes the polarizer with probability $\langle v^{*}_{k}|\rho_{\rm{in}} |v^{*}_{k}\rangle$ and then results in, due to the projection at the polarizer, the polarization state $|H\rangle$. Once the input single-photon qubit has passed the polarizer (i.e., projection measurement has occurred), the corresponding state $|v_{k}\rangle$ is prepared from the state $|H\rangle$ using another set of waveplates, see Fig.~\ref{fig:setup}(b). Note that, since we employ the triggered single-photon source for encoding $\rho$, only the coincidence event between the signal detector and the trigger detector is meaningful. The coincidence event can occur only when the signal photon has passed through the polarizer P in the measurement stage in  Fig.~\ref{fig:setup}(b). Thus, when the triggered single-photon source is used, the detector and the single-photon source in the measurement and the state preparation stages, respectively, shown in Fig.~\ref{fig:setup}(b) become redundant and can be removed altogether as we have done in our work.

\begin{figure}[t]
\includegraphics[width=3.4in]{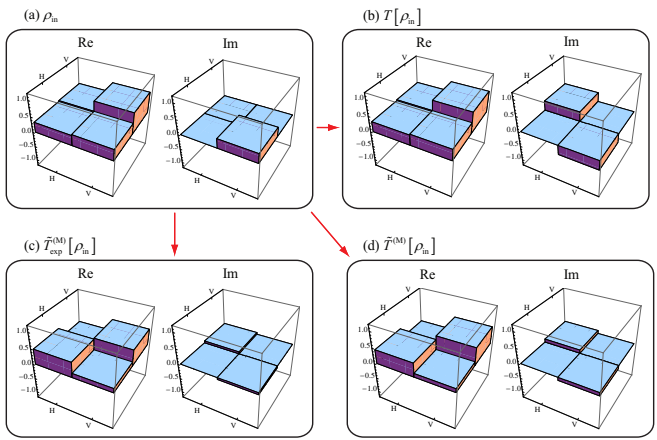}
\caption{(a) QST of the qubit state $\rho_{\rm{in}}$ in eq.~(\ref{eq:rhoin}). (b) The state after transpose, $T[\rho_{\rm{in}}]$. (c) QST of the qubit state after the experimental $\widetilde{T}_{\rm{exp}}^{(M)}$ operation. (d) The state after the ideal approximate transpose. Note that the fidelity between (b) and (d) is $F= 2/3$, as it is shown in eq.~(\ref{eq:2/3}). The Uhlmann's fidelity between (c) and (d) is, $F\approx 0.996$.}
\label{fig:QST}
\end{figure}

Finally, for a given state $\rho_{\rm{in}}$, the SPA to the transpose $ \widetilde{T}_{\rm{exp}}^{(M)} [\rho_{\rm{in}}]$ is constructed by the probabilistic sum of four equally-weighted $\widetilde{T}^{(M)}_k$ operations. (In other words, this means that the four paths in Fig.~\ref{fig:setup}(a) are non-interfering.) The resulting state is identified by the quantum state tomography (QST) and is then quantified by comparing to the ideal case, $\widetilde{T}^{(M)} [\rho_{\rm{in}}]$. The experimentally implemented transpose operation is also analyzed by the quantum process tomography (QPT) that identifies the performed operation. In the experiment, the measurement duration in each setting was 1 s and the measurement was repeated three times. QST and QPT results were obtained using the maximum likelihood estimation.


To apply the approximate transpose operation, we consider an arbitrary polarization state, $|\psi\rangle = (|H\rangle + (1+i)|V\rangle)/\sqrt{3}$, and the experimentally prepared one is identified by the QST, see Fig. \ref{fig:QST} (a):
\bea \rho_{\rm{in}} = \left( {\begin{array}{*{20}c}
   0.322 & 0.352 - 0.307i  \\
   0.352 + 0.307 i & 0.678  \\
\end{array}} \right) \approx |\psi\rangle\langle\psi|. \label{eq:rhoin}\eea
The transposed state can easily be computed as, $T[\rho_{\rm{in}}] = \rho_{\rm{in}}^{T}$, which is shown in Fig.~\ref{fig:QST} (b). The experimental result for the measurement-based SPA scheme depicted in Fig.~\ref{fig:setup} is then shown in Fig.~\ref{fig:QST} (c). Assuming the ideal application of the approximate transpose, the resulting state would be $\widetilde{T}^{(M)}[\rho_{\rm{in}}]$ from eq.~(\ref{eq:spat}) and is shown in Fig.~\ref{fig:QST} (d). Using the Uhlmann's fidelity, the experimental result can be quantified in terms of its similarity with the ideal one, $F( \widetilde{T}^{(M)}[\rho_{\rm{in}}], \widetilde{T}_{\rm{exp}}^{(M)}[\rho_{\rm{in}}]) \approx 0.996$. We have also repeated the approximate transpose to a few more states, and obtained similar values of the fidelity.

\begin{figure}[t]
\includegraphics[width=3in]{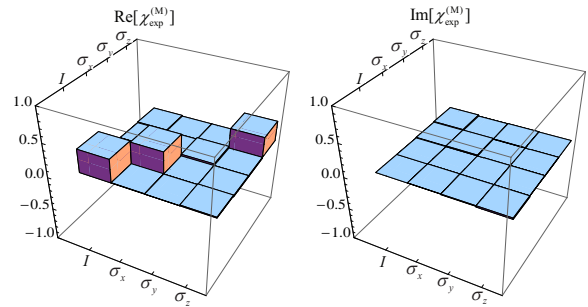}
\caption{The $\chi$ matrix for the operation $\widetilde{T}_{\rm{exp}}^{(M)}$ is obtained from the QPT. The average fidelity between the performed and the ideal operations is, $F_{\rm{ave}} (\widetilde{T}_{\rm{exp}}^{(M)},\widetilde{T}^{(M)}) \approx 0.999$.}\label{fig:QPT}
\end{figure}


The actual operation that has been performed in experiment can be found by the QPT. In this way, the similarity between designed and performed quantum operations can be estimated. For the QPT, note that the Pauli matrices $(\sigma_{0}(=\Id), \sigma_{x}, \sigma_{y}, \sigma_{z})$ span the operator space of single qubit operations. Hence, a quantum process $\E$ of a single qubit can generally be expressed as, $\E(\rho_{\rm{in}}) = \sum_{m,n} \chi_{mn} \sigma_{m} \rho_{\rm{in}} \sigma_{n}^{\dagger}$, where it is the matrix $\chi_{mn}$ that gives the complete characterization of the operation $\E$. For the ideal operation $\widetilde{T}^{(M)}$, the corresponding $\chi$ matrix is found to be, $\chi(\widetilde{T})=diag[1/3,1/3,0,1/3]$. After the QPT, the $\chi$ matrix of the performed operation $\widetilde{T}_{\rm{exp}}^{(M)}$ has been constructed, and is shown in Fig.~\ref{fig:QPT}. The average fidelity between the designed and the performed operations is exploited to compare two channels, $F_{\rm{ave}}(\widetilde{T}^{(M)}, \widetilde{T}_{\rm{exp}}^{(M)})  \approx 0.999$ \cite{ref:avefi}.

So far, we have shown an experimental implementation of the universal transpose using the measurement-based SPA scheme that can give the maximal fidelity. The scheme can then be used as a building block for further applications of approximate positive maps such as entanglement detection \cite{ref:HorodeckiEkert,ref:spalocc}. We emphasize that the presented scheme is practical within the present-day technology as only linear optics are used. There has been an earlier result reported in Ref.~\cite{ref:Sciarrino}, where the universal transpose is implemented by a random unitary channel. As it was pointed out in the original proposal in Ref. \cite{ref:HorodeckiEkert}, the scheme based on a unitary channel requires quantum memory at the final step for the spectrum estimation. This contrasts to the measurement-based SPA scheme where no quantum memory is required \cite{ref:Korbicz}.

\begin{figure}[t]
\includegraphics[width=3in]{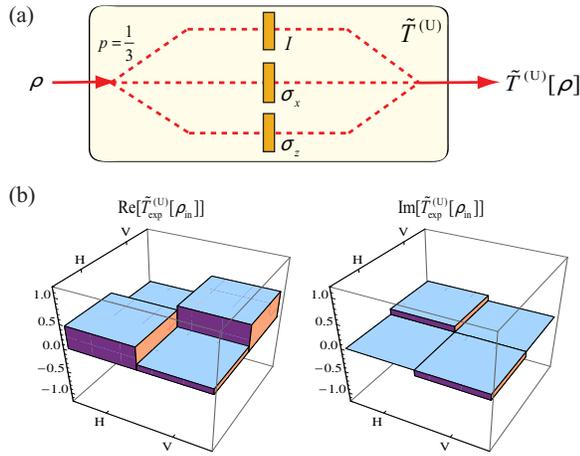}
\caption{(a) The scheme $\widetilde{T}^{(U)}$ is based on random applications of three unitaries, $I$, $\sigma_{x}$, and $\sigma_{z}$ on a single qubit. The unitaries $\sigma_k$ can be realized with half-wave plates. (b) The resulting state after applying $\widetilde{T}^{(U)}$ to the input state $\rho_{\rm{in}}$ in eq.~(\ref{eq:rhoin}) is identified by the QST. The fidelity with the ideal one is, $F(\widetilde{T}^{(U)}[\rho_{\rm{in}}], \widetilde{T}_{\rm{exp}}^{(U)}[\rho_{\rm{in}}]) \approx 0.999$.
}\label{fig:unitary}
\end{figure}

If we focus only on the implementation of the universal transpose, the unitary-based scheme has an advantage over the measurement-based SPA scheme in that lesser numbers of optical elements are needed which helps improving the fidelity. To fairly compare the two schemes, we have also implemented the unitary-based scheme for the transpose: the unitary channel can be found from the Jamiolkowski state in eq.~(\ref{eq:jamistate}), $\rho_{\widetilde{T}}= [\Id\otimes\widetilde{T}](|\phi^{+}\rangle\langle\phi^{+}|) = \frac{1}{3} \sum_{i=0,x,z} (\Id \otimes \sigma_i) |\phi^{+}\rangle\langle\phi^{+}| (\Id \otimes \sigma_i)$. Hence, using the relation in Ref.~\cite{ref:Jamiolkowski} one can derive that, $ \widetilde{T}^{(U)}[\rho_{\rm{in}}] = \frac{1}{3} \sum_{i=0,x,z} \sigma_{i} \rho_{\rm{in}} \sigma_{i}$, where the superscript $(U)$ means that the scheme is unitary-based \cite{ref:Sciarrino}. In Fig.~\ref{fig:unitary}, the experiment results are shown for the state in eq.~(\ref{eq:rhoin}). The  average fidelity between the experimentally obtained QPT $\chi$ matrix of the operation $\widetilde{ T }_{\rm{exp}}^{(U)}$ and the ideal operation is $F_{\rm{ave}}(\widetilde{T}^{(U)}, \widetilde{T}_{\rm{exp}}^{(U)})  \approx 0.999$.

Let us also comment on the the relation of our work with other implementations of approximate anti-unitary operations. The universal-not (UNOT) operation that flips unknown quantum state is anti-unitary and expressed as the product of the Pauli matrix $\sigma_{y}$ and the transpose \cite{ref:optimal}. In Ref.~\cite{ref:DeMartini}, the optimal approximate UNOT operation has been realized in experiment by making use of the anti-cloning process that appears in the ancillary system of the $1\rightarrow 2$ symmetric universal quantum cloning \cite{ref:anticl}.  Since the involved quantum cloning process is an entangling operation that necessarily requires the CNOT operation in the scheme, nonlinearity that generally gives rise to a lower efficiency was needed to enact interactions between the photons.

In conclusion, we have shown that an optimal approximate transpose operation can be realized with measurement and preparation of quantum states. To the best of our knowledge, this is the first proof-of-principle demonstration that shows SPAs to positive maps can be implemented with present-day technologies (i.e., without requiring quantum memory and collective measurement). We believe that our work sheds new light on the practical implementation of SPAs to positive maps, which is closely related to experimental detection of entanglement.


This work was supported by  National Research Foundation of Korea (2009-0070668, 2009-0084473, and  KRF-2008-313-C00185) and the Ministry of Knowledge and Economy of Korea through the Ultrafast Quantum Beam Facility Program. YSK acknowledges the support of the Korea Research Foundation (KRF-2007-511-C00004).

\vspace*{-0.3cm}


\end{document}